\documentclass[jkps,preprint,fleqn,showpacs,showkeys]{revtex4}
\usepackage{graphicx}
\usepackage{amssymb}
\usepackage{amsmath}
\usepackage{bm}
\begin{document}
\setcounter{page}{1}
\title[]{Delta Scorpii 2011 periastron: worldwide observational campaign and preliminary photometric analysis}
\author{Costantino \surname{Sigismondi}}

\email{sigismondi@icra.it}
\thanks{AAVSO observer's code SGQ}
\affiliation{Sapienza University of Rome and ICRA, International Center for Relativistic Astrophysics}

\date[]{Received 5 July 2011 h. 20 UT}

\begin{abstract}
Delta Scorpii is a double giant Be star in the forefront of the Scorpio, well visible to the naked eye, being normally of magnitude 2.3. In the year 2000 its luminosity rose up suddenly to the magnitude 1.6, changing the usual aspect of the constellation of Scorpio. This phenomenon has been associated to the close periastron of the companion, orbiting on a elongate ellipse with a period of about 11 years. The periastron, on basis of high precision astrometry, is expected to occur in the first decade of July 2011, and the second star of the system is approaching the atmosphere of the primary, whose circumstellar disk has a H-alpha diameter of 5 milliarcsec, comparable with the periastron distance. The preliminary results of a photometric campaign, here presented in the very days of the periastron, show an irregular behavior of the star's luminosity, which can reflect some shocks between material around the two stars. The small luminosity increasement detected in the observation of 5 of July 2011 at 20 UT may suggest that the periastron phenomena are now going to start.
\end{abstract}



\maketitle

\section{INTRODUCTION: BRIGHT STARS PHOTOMETRY}

The first star recognized as variable was Mira Ceti in 1636 by Holwarda,
even if it was discovered by David Fabricius in 1595 and seen again in 1607, always in the vicinity of Jupiter.
The antique observers were not interested to stellar variability,\cite{Thomas} and the stars were considered as $"fixae"$ ($fixed$ in Latin) either regarding to their position and implicitly also regarding to their luminosity.\cite{SigismondiAAVSO}
Algol, for example, was already known by the ancient Arabs who gave to it this name, meaning "the Devil's eye", but its regular variability was correctly explained only by John Goodricke in 1784 with the eclipse mechanism in a double system with the orbital plane parallel to the line of sight.\cite{Goodricke}

Maxima and minima of Mira variables have some magnitudes of difference, while smaller variations occur for the Cefeids stars.
Irregular stars can change of a whole magnitude or so, without regularity, and there are also very bright stars among them. The two supergiant M-type stars Betelgeuse,\cite{Ravi} and
Antares\cite{Simonsen} are famous examples, and also $\delta$ Scorpii. These are three stars visible from both emispheres, which deserve a more systematic and accurate monitoring. 

The naked eye dataset for such bright stars [in fig. 1 and 2 there is the example of $\delta$ Scorpii] present intrinsical scatters comparable with the real variations to be studied, so it is complicate to identify real fluctuations of the stellar luminosity with respect to measurement's errors.
Better results are obtained selecting the photometric band V as in fig. 2.

\begin{figure}
\includegraphics[width=17.0cm]{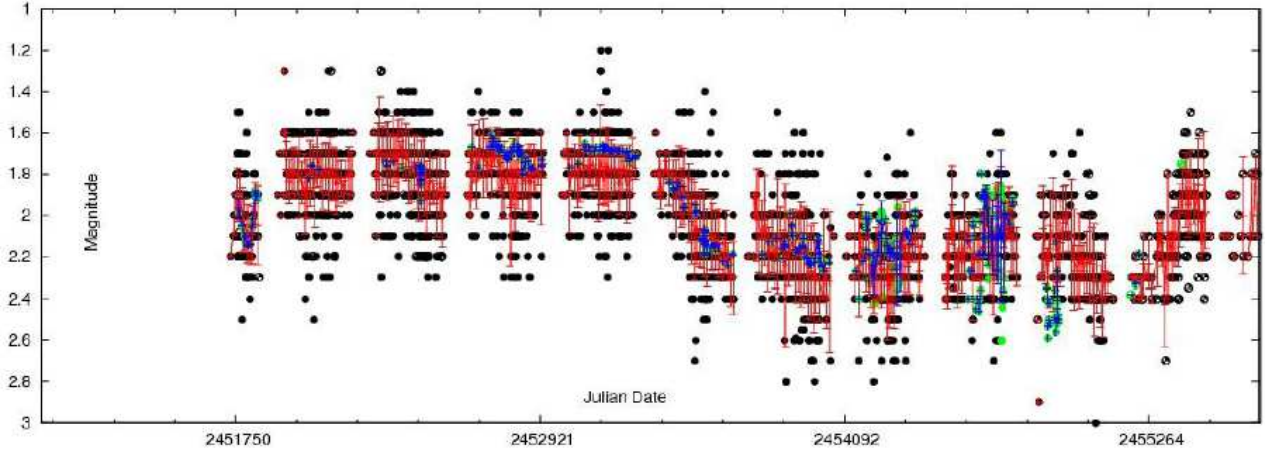}
\caption{The observations of $\delta$ Scorpii with naked eye since the year 2000. The real amplitude of luminosity variations is overlapped with measurement's uncertainties at the naked eye. AAVSO data.}
\label{fig1}
\end{figure}

\begin{figure}
\includegraphics[width=17.0cm]{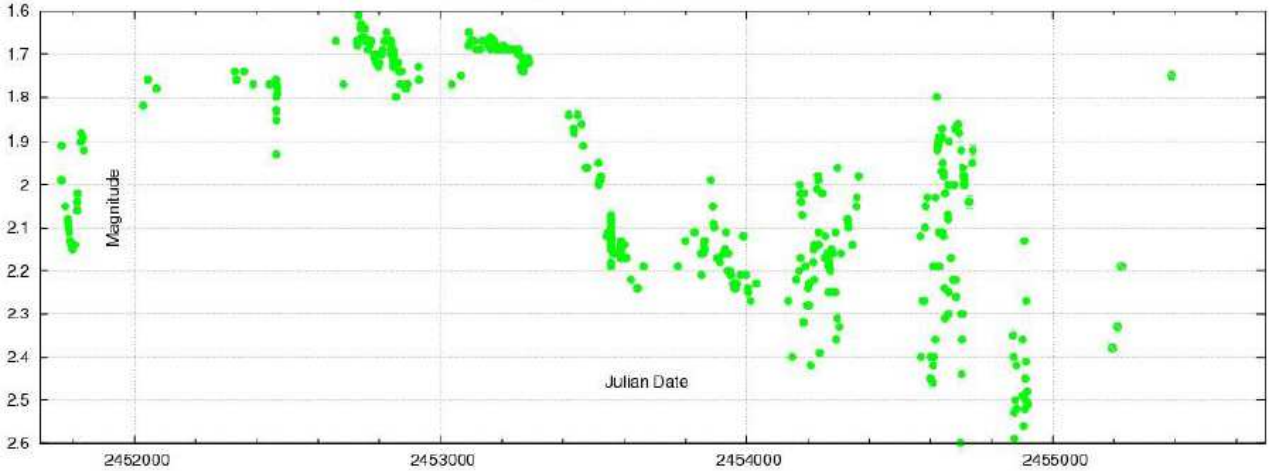}
\caption{The observations of $\delta$ Scorpii in V band since the year 2000. These measurements show luminosity variations probably intrinsic to this double system, but in the last 4 years the data points are rather spread, probably loosing the physical information under the measurement's error. 
AAVSO data.}
\label{fig1}
\end{figure}

\section{THE AAVSO: 100 YEARS OF CITIZEN ASTRONOMY}

In the year 2011 the AAVSO, American Association of Variable Stars Observers, complete its first 100 years of activity.
The slogan chosen for this centennial year is "100 years of citizen astronomy"\cite{aavso}
and reflects the philosophy of that association: to parallel the professional astronomers in monitoring of the variable stars.
About 20.5 millions observations are stored in the AAVSO database, which contains naked eye observations as well as CCD, photometric and other kinds of observations, the same database is also opened to the scientific research of professional astronomers, who often request AAVSO observers to provide accurate observations of target stars in the days before and after their observational campaign.

\section{DELTA SCORPII 2011 PERIASTRON}

The double system of $\delta$ Scorpii is undergoing a very close periastron in the first decade of July 2011 according to the predictions of two research teams.\cite{Tyc}, \cite{Fizeau}
The discovery of the variability of Delta Scorpii was done by Sebastian Otero,\cite{Otero} a spanish amateur astronomer, in May 2000. It has been hypotized that this luminosity variation was triggered by the periastron of the companion stars, orbiting in a very eccentric orbit with $e=0.97$. The periastron occurred in September 2000, with the star invisible because immersed in the light of the Sun. This year, 2011, there is the possibility to observe the periastron with the star visibile during most part of the night.
Accurate measurements of astrometric quality, carried independently by the two teams above mentioned, show that the 2011 periastron occurs with the two stellar atmospheres which can get in contact each other.
The observational campaign of the current periastron has been recommended on specifical websites.\cite{sigiweb},\cite{Mirosh}    

\section{DIFFERENTIAL PHOTOMETRY WITH NAKED EYE}

The technique of making differential photometry, comparing the luminosity of some reference stars close the the target star, overcome the problem of the absolute calibration of detectors, and of the atmospheric extinction rapidly varying when the stars approach the horizon. 

The problem of monitoring bright irregular variable stars is the lack of reference with brighter stars.

Furthermore, it is important that the reference stars are not too much fainter than the target star, in order to avoid the saturation of the target star itself in the images.

Using differential photometry techniques avoids rather well the problem of differential atmospheric extinction, if the stars are enough close each other.
In the case of $\delta$ Scorpii the reference stars are within $10^o$ of angular distance, and at the same altitude above the horizon (about $20^o$ for Antares and $\delta$) or $5^o$ of altitude's difference ($\beta$ above $\delta$). The correction for $\delta$ compared with $\beta$ in this case is less than +0.1 magnitude, while no correction is needed in the comparison with Antares, of which I consider its V magnitude.

Another possibility, for naked eye observations, is to compare the difference of luminosity between $\delta$ and Antares and between $\delta$ and $beta$ with the difference of luminosity of other selected pairs of bright stars visible in the same time.

An example of this method is to use the pair $\alpha$ Lyrae, Vega, and $\alpha$ Aquilae, Altair 

and the pair Vega and $\alpha$ Cygni, Deneb: 

these stars are all enough bright to be seen under city skyes, even if heavily polluted by the urban luminosity glow. 

For the Southern emisphere the pair $\alpha$ and $\beta$ Centauri, has been selected to verify the "amplitude" between $\alpha, \delta$ and $\beta$ Scorpii using the naked eye.

In general for differential photometry it is always better to select pairs with similar colour index with respect to the target star.

For example in the case of $\delta$ Scorpii, I have selected the stars $\beta$ and $\pi$ Scorpii for performing the differential photometry, because of their colour index, very similar to $\delta$ Scorpii:  $|B-V|\le 0.2$.

In the naked eye observations, the need of comparing $\delta$ Scorpii, a Be-type star with $\alpha$, Antares, which is M-type star, even irregular, can introduce some errors. I have considered the V magnitude of Antares as published on SIMBAD catalogue\cite{simbad} in the data analysis.

All naked eye observation have been overlapped with CMOS imaging, only from the airplane above the Atlantic Ocean it was not possible to make a photo.

\section{DIFFERENTIAL PHOTOMETRY WITH LARGE FIELD DIGITAL PHOTO}

A commercial camcorder with CMOS detector has been used for the photos. It is a SANYO CG9 8-bit camcorder at 800 ISO and $3\times$ magnification; the time of pose is automatically set to 0.5 s and allows to get the $4^{th}$ magnitude: this field of view spans approximately twice the angular distance between $\alpha$ and $\delta$ Scorpii. The camera lens is 1.6 cm of diameter.

For each observative session several photos of the stellar field were taken, and the better images were selected and processed with IRIS v. 5.59 software.\cite{IRIS} 
IRIS can perform either "Aperture Photometry" and "RGB Separation" in order to select the Green (G) band, considered the most close to the V band of Johnson photometric system.\cite{Fiocco}

Owing to the low signal-to-noise ratio of several images, and to the presence of stray lights polluting the image, due to the city lights, and also to the inhomogeneities of the sky glow over $10^o$, the aperture photometry has not given very good results.

The method with better results has been the one of "best 3 pixels".

This method consists into selecting the 3 more luminous pixels of a given star in the G channel. Each pixel has up to 255 counts, and the luminous intensity of a star is represented by the sum of the three pixels' values.
The logarithm of the ratios between the counts for $\delta$ with $\beta$ and $\pi$ Scorpii has been multiplied by 2.5 to recover the difference in magnitudes with the two reference stars. 
The errorbars associated to these measurements are obtained by the half-differences of the estimated magnitudes calculated using respectively $\beta$ or $\pi$ Scorpii as a reference stars.     

For about half of the observations the star was at the zenith in Rio de Janeiro. The others observations are made in Rio or in Italy with the star lower in the Southern sky. The images are affected by sky glow, because they were obtaiened in an urban environment.
When the sky glow and the effect of the stray lights affected evidently the images, the visual data -obtained with naked eye- have been preferred (see fig. 3).

The scintillation is a problem for the observations near the horizon.

The maximum errorbar in this case has been estimated with the formula\cite{AAVSOman}

$\Delta M_{scint} = \frac{0.09\cdot A^{7/4}}{D^{2/3}\cdot \sqrt{2t}}$

with D= 2 cm, the maximum airmass value A=3 and t=0.5 s,
$\Delta M_{scint} = 0.4$ magnitudes.

The scintillation and the anisotropies of the sky glow or of thin clouds over $10^o$, when the star was higher in the sky (with $A\sim 1$), produced the main source of uncertainty from the digital images.
To the naked eye, all these effects are reduced because the evaluation and the comparisons of the luminosities are done over about 5-10 minutes each time.

\begin{figure}
\includegraphics[width=17.0cm]{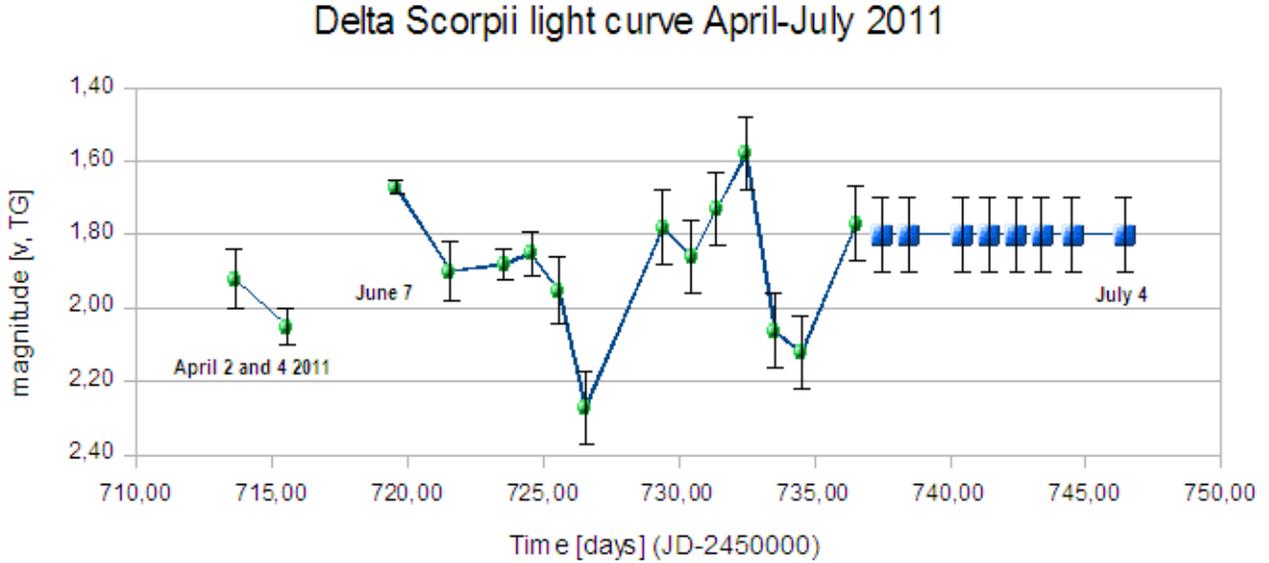}
\caption{The series of the observations of Delta Scorpii, since 2 April 2011. The observations are published and constantly updated in the AAVSO website http://www.aavso.org/ql/results?auid=000-BBW-851\&startjd=2455653.5 with the acronym SGQ, associated since 1999 to myself as a contributing observer. 
The green balls are the data obtained with CMOS image's analysis in the green channel TG, the blue squares are visual observations, treatable as V mangnitudes. The visual data of the last 10 days are obtained with the stars affected by the sky glow, they appear more stable than the corresponding CMOS images, and have represented in this image. The last visual observation, not represented in this figure, is $M_V 1.65\pm 0.1$ of July 5 20 UT.}
\label{fig1}
\end{figure}

\section{CONCLUSIONS}
The data here presented for the $\delta$ Scorpii 2011 periastron campaign are still preliminary. 
The variations of luminosity plotted in fig. 3 can suggesting the existence of shock waves between the approaching stellar exospheres.
The lack of significant luminosity variations in the last week can signify that the close periastron phenomena are going to occur in the next days, or that the phenomenon is not detectable as an overall change of luminosity of the whole system.

The fluctuations of the digital TG luminosity detected can be done also to the influence of stray lights or inomogeneous sky glow or thin high clouds back illuminated by the city lights.
 
In order to avoid this risk several images of the same observing session have been examined. When the results of the estimated magnitude for $\delta$ Scorpii were consistent among different images the effect was considered as intrinsical to the star and plotted in the fig. 3.

To avoid the anti-blooming gate ABG\cite{AAVSOman} of the CMOS detector of the camera it is better to limit the pixel saturation to the $50\%$ level. The ABG effect destroy the linearity of the detector, and the reliability of the photometry, that's why in some cases the sensitivity of the camera was set to 400 ISO, in order to avoid pixel's saturation. 

The recommended procedure, in order to collect reliable data, is the simultaneous observation with naked eye and with photo, and among naked eye procedure it is also to recommend the selection of pairs of bright stars separated by a known amount of magnitudes, in order to use them as standard "rulers" of magnitude's difference.

With observations with the sky still bright, or with the eye not accomodated to the darkness, the procedure exploiting the eye accomodation time and the physiology of cones and rods\cite{sigi2000} can be also used, once it has been calibrated with opportune pairs of reference stars.
 
\begin{acknowledgments}
Thanks to Claudio Lopresti for the invitation to the XXV National meeting on Variable Stars Observations in La Spezia -Italy on 14 May 2011, where this subject was discussed first, to the ICRANet of Pescara-Italy for welcoming this topic in the Italian-Korean XII meeting of Relativistic Astrophysics from 4 to 8 July 2011, which is going on during the periastron of $\delta$ Scorpii; and to Francesca De Rosa for her assistance and patient collaboration during the observations of the star and the photos. A FACC grant is also acknowledged. 
\end{acknowledgments}

\end{document}